\documentclass[aps,prl,twocolumn,amsmath,amssymb,10pt]{revtex4-1}

\usepackage{graphicx}
\usepackage{dcolumn}
\usepackage{bm}
\usepackage{xcolor}
\usepackage[flushleft]{threeparttable}

\begin{document}

\title{Temperature-dependent rotationally inelastic collisions of OH$^-$ and He}

\author{Eric S. Endres$^{1}$}
\author{Steve Ndengu\'e$^{2,3}$}
\author{Olga Lakhmanskaya$^1$}
\author{Seunghyun Lee$^1$}
\author{Francesco A. Gianturco$^1$}
\author{Richard Dawes$^2$}
\author{Roland Wester$^1$}
\email{roland.wester@uibk.ac.at}

\affiliation{$^1$Institut f\"ur Ionenphysik und Angewandte Physik, Universit\"at Innsbruck, Technikerstrasse 25/3, 6020 Innsbruck, Austria}

\affiliation{$^2$Department of Chemistry, Missouri University of Science and Technology, 65409 Rolla, Missouri, United States.}

\affiliation{$^3$ICTP-East African Institute for Fundamental Research, University of Rwanda, Kigali, Rwanda}


\date{\today}

\begin{abstract}
We have studied the fundamental rotational relaxation and excitation collision of OH$^-$ $J=0 \leftrightarrow 1$ with helium at different collision energies. Using state-selected photodetachment in a cryogenic ion trap, the collisional excitation of the first excited rotational state of OH$^-$ has been investigated and absolute inelastic collision rate coefficients have been extracted for collision temperatures between 20 and 35\,K. The rates are compared with accurate quantum scattering calculations for three different potential energy surfaces. Good agreement is found within the experimental accuracy, but the experimental trend of increasing collision rates with temperature is only in part reflected in the calculations.
\end{abstract}

\maketitle

Experimental advances in the preparation of ensembles of cold and ultracold molecules have enriched the field of cold chemistry with studies of inelastic and reactive scattering at low temperatures \cite{Doyle2004:epjd, Carr2009:njp, Dulieu2011:pccp}. Low temperatures enable rotational state control for molecules and reveal quantum properties of collisions, such as orbiting and shape resonances \cite{Koch2019:rmp,  Klein2016:np}. Furthermore, control over molecular states has potential applications in fundamental precision studies \cite{Safronova2018:rmp, Hudson2002:prl, Hudson2006:prl, Schiller2014:prl, Germann2014:np}, quantum information processing \cite{Andre2006:np,DeMille2002:prl} and many-body physics \cite{Koch2019:rmp, Moses2016:np}.

Cooling molecules, unlike atoms, implies freezing of the internal degrees of freedom. For molecular ions the general way to relax all degrees of freedom is collisional dissipation of energy using buffer gas cooling \cite{Nguyen2011:njp, Hudson2016:epj, Willitsch2012:irpc, Wester2009:jpb}. In standard cryostats, temperatures are limited to above 3\,K. Lower temperatures may be reached in hybrid atom-ion-traps, where a magneto-optical trap for ultracold atoms is superimposed with an ion trap \cite{Hudson2016:epj}. Understanding the rotational quenching kinetics and, more specifically, state-specific inelastic collision rates is necessary to be able to control and manipulate the internal state population of trapped molecular ions.

Inelastic collision studies are also of great relevance for astrophysics, to model relaxation kinetics in the early universe or to describe molecular excitation levels in interstellar molecular clouds \cite{Lique2019:iopa, Smith2011:araa}. This is particularly important when local thermodynamic equilibrium (LTE) cannot be assumed. Rate coefficients for rotationally inelastic collisions are needed to bring the predicted line intensities into agreement with astronomical observations or to quantitatively correlate deuterium to hydrogen abundance ratios with the conditions in astrophysical environments \cite{Gerlich2002a:pss}. Rotational state control is also needed for state-selected ion-molecule reaction studies in order to better understand the gas phase ion chemistry that dominates in cold interstellar clouds \cite{Larsson2012:rpp}.

Given their importance, numerous theoretical calculations provide rate coefficients for inelastic rotational state-changing collisions of molecular ions, e.g. for NO$^+$ \cite{Denis-Alpizar2015:mnras}, C$_6$H$^-$ \cite{Walker2016:mnras}, H$_2^+$ \cite{Hernandez-Vera2017:jcp}, or C$_2$H$^-$ and C$_2$N$^-$ \cite{Franz2020:jcp}. However, only few experiments have been able to provide absolute rate coefficients, in particular at low temperatures. Schlemmer {\it et al.\ } have investigated rotational cooling of N$_2^+$ colliding with Argon using a laser-induced reaction \cite{Schlemmer1999:ijms}. Hansen {\it et al.\ } have studied rotational cooling, but did not extract absolute rate coefficients \cite{Hansen2014:n}. In Ref.\ \cite{Hauser2015:np} we have introduced a novel scheme to measure inelastic collision rate coefficients via state-specific photodetachment, which is applicable to negatively charged molecules. With this we have obtained the inelastic collision rate coefficients that link the two lowest rotational states of OH$^-$ and OD$^-$ anions in collisions with He.

In this letter we present experimental results for the temperature dependence of the rotationally inelastic collision rate coefficient of OH$^-$($^1\Sigma^+$) colliding with He using state-specific photodetachment \cite{Hauser2015:np}. We use the experimental results to benchmark three different quantum scattering calculations. OH$^-$ is particularly well suited for such studies due to the simple rotational structure in its $^1\Sigma^+$ ground rotational state, the large rotational constant of 562\,GHz \cite{Jusko2014:prl}) and the well studied properties of near threshold photodetachment \cite{Schulz1982:jcp,Goldfarb2005:jcp,Engelking1984:pra,Smith1997:pra}.

A detailed description of the experimental set up can be found elsewhere \cite{Otto2013:pccp,endres2017:jms}. Briefly, OH$^-$ anions are produced in a plasma discharge of a helium water mixture and are loaded in a 22-pole radiofrequency ion trap after mass selection. The trap is filled with He buffer gas, which collisionally thermalizes the kinetic ion temperatures and internal degrees of freedom. Buffer gas temperature was varied from 9\,K to 30\,K. The lifetime of the ions typically exceeds thousands of seconds, and thus, does not affect our measurements. The employed helium densities range between $2.7 \cdot 10^{11}$\,cm$^{-3}$  and $7 \cdot 10^{11}$\,cm$^{-3}$. After a short thermalization period, the photodetachment laser is admitted into the trap with powers between 70\,mW and 210\,mW.  This initiated the ion losses with rates ranging from 0.2\,s$^{-1}$ to about 2\,s$^{-1}$. The interaction time with the laser was regulated using a self-built mechanical laser shutter.

In the ion trap, OH$^-$ ions undergo collisions with the buffer gas, which couples the rotational states of OH$^-$ via inelastic collisions and establishes a Boltzmann distribution of the rotational level population at a given temperature. The admission of the photodetachment (PD) laser to the trap initiates the loss of ions from the excited rotational state $J=1$ and higher with a rate $\gamma_{PD}$ (see inset in Fig.\ 1). Due to the presence of inelastic collisional coupling of the rotational state $J=0$ and $J=1$ through the thermal collision rates $\gamma_{01}$ and $\gamma_{10}$, the first excited rotational state gets repopulated. Thus the absolute number of anions in the ground state decays as well. For low enough buffer gas density and high enough laser power this coupling becomes insufficient to maintain the original thermal distribution of the rotational state populations. This leads to a non-linear dependence of the ion losses on laser power.

\begin{figure}
\includegraphics[width=\columnwidth]{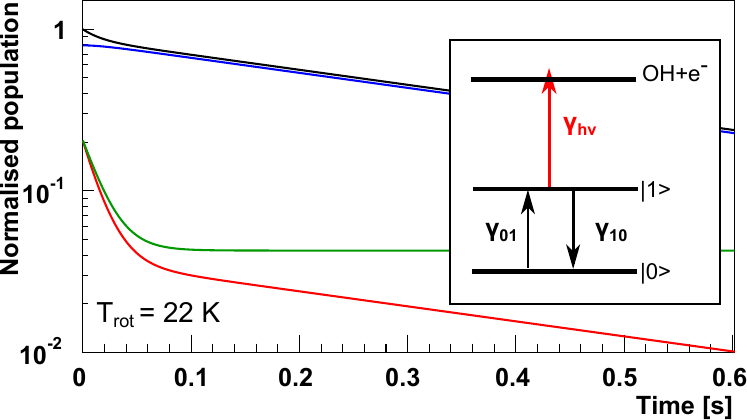}
\caption{Time evolution of the rotational state population of OH$^-$ at 22\,K rotational temperature. The blue line represents the ground state population $N_0$ and the red line the excited state population $N_1$. The black and green lines show the total population $N_0+N_1$ and the fraction $N_1/(N_0+N_1)$, respectively. The graph was simulated with the parameters $\gamma_{PD}=50\,s^{-1}$ and $\gamma_{10}=10\,s^{-1}$.}
\label{fig:pop_vs_t}
\end{figure}

The time evolution of the rotation state population is calculated by solving the coupled rate equations
\begin{equation}
    \frac{d}{dt}\left(\begin{array}{c}
N_{0}\\
N_{1}
\end{array}\right)=\left(\begin{array}{cc}
-\gamma_{01} & \gamma_{10}\\
\gamma_{01} & -\left(\gamma_{10}+\gamma_{PD}\right)
\end{array}\right)\left(\begin{array}{c}
N_{0}\\
N_{1}
\end{array}\right).
\label{eq:system_dif_eq}
\end{equation}
Here OH$^-$ is approximated as a two-level system with populations $N_0$ (ground state) and $N_1$ (excited state). This is a good approximation in the considered temperature range due to the large rotational constant of OH$^-$. The sum of the two equations provides the instantaneous loss rate of the ions
\begin{equation}
\gamma_L(t) = \frac{1}{(N_{0}+N_{1})}\frac{d}{dt}\left(N_{0} +N_{1}\right)=-\frac{N_{1}}{N_{0}+N_{1}} \gamma_{PD}
\label{eq: loss rate}
\end{equation}
The rate is time-dependent as the relative population of the excited state changes with time, which leads in general to a non-exponential decay of the trapped ion number. This is shown in Fig.\ 1, which presents a solution of Eq.\ (\ref{eq:system_dif_eq}) for the case of a high photodetachment rate compared to the inelastic relaxation rate. The figure shows the general result of a fast initial relaxation of the relative excited state population $\frac{N_1}{N_0+N_1}$ on a time scale $\tau \sim 1/(\gamma_{01}+\gamma_{10}+\gamma_{PD})$ after which it stays constant. For times $t \gg \tau$ one can simplify the solution of Eq.\ (\ref{eq: loss rate}) to an exponential decay with a constant loss rate 
\begin{equation}
\begin{split}
& \gamma_L(t\rightarrow \infty) = \\
& \frac{2\gamma_{01} \gamma_{PD}}{\gamma_{01}+\gamma_{10}+\gamma_{PD}+\sqrt{(\gamma_{01}+\gamma_{10}+\gamma_{PD})^2-4\gamma_{01}\gamma_{PD}}} \end{split}
\label{eq:fit}
\end{equation}
When the collisional coupling of rotational states is strong compared to the photodetachment rate ($\gamma_{PD}\ll \gamma_{01},\,\gamma_{10}$), the ratio $\frac{N_1}{N_0+N_1}$ stays constant. As a result $\gamma_L$ increases linearly with $\gamma_{PD}$ (see. eq.\ \ref{eq: loss rate}).

The thermal excitation and de-excitation rate coefficients are coupled by detailed balance,  $\gamma_{01}/\gamma_{10}=g_1/g_0 \, \exp\left(-\Delta E/k_B T_{rot}\right)$, to the temperature $T_{rot}$ that describes the rotational population. This temperature also represents the collision temperature in the center of mass frame of the OH$^-$/He system\cite{Wester2009:jpb}. It can be determined independently at a low photodetachment rate following the rotational thermometry scheme described in Ref.\ \cite{Otto2013:pccp}.

The photodetachment rate $\gamma_{PD}$ is proportional to the laser power admitted into the trap. It could in principle be determined using an absolute photodetachment cross section measurement \cite{Hlavenka2009:jcp,Best2011:aj}. However, here we are only interested in the linear dependence of $\gamma_{PD}$ on the measured laser power $P$ and therefore use a free parameter $\alpha$ linking $\gamma_{PD} = \alpha P$ \cite{Gianturco2018:fd}. As a consequence two free parameters remain in Eq. (\ref{eq:fit}), $\gamma_{10}$ and $\alpha$, that need to be fitted to the experimental data.

\begin{figure}[tb]
\includegraphics[width=\columnwidth]{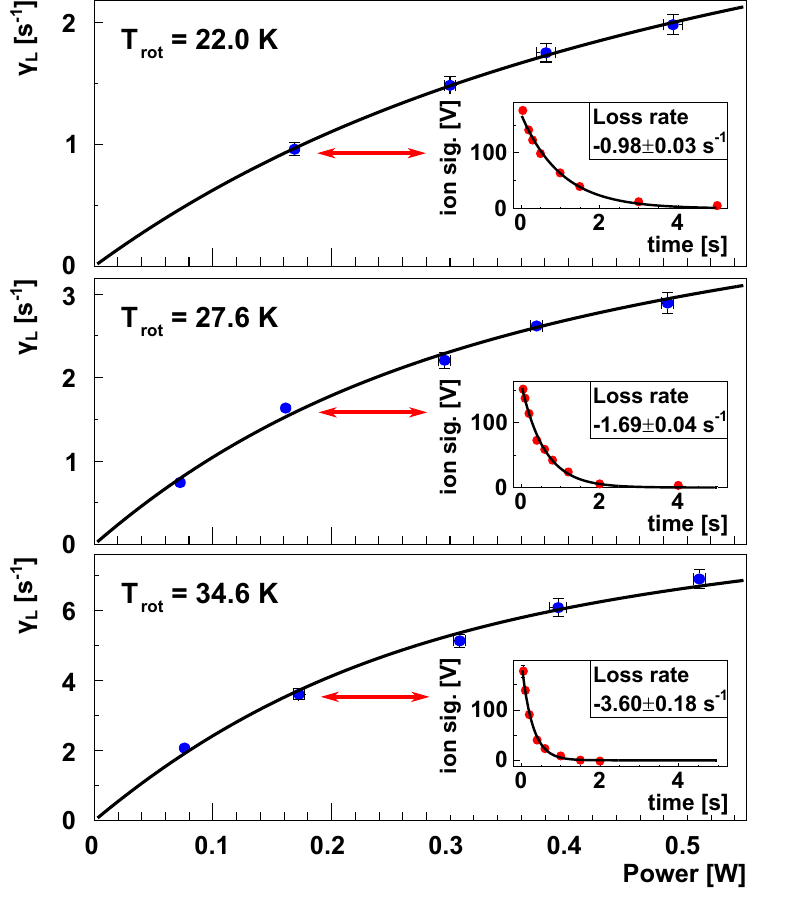}
\caption{Power dependent loss rate (blue data points) fitted by the solution to the two level rate equation system (black line). The insets show selected individual loss rate measurements. The data have been obtained at trap temperatures that yield the specified rotational temperatures 9\,K, 20\,K and 30\,K \cite{endres2017:jms}. Fit parameters and employed buffer gas densities are reported in Table \ref{tab:summary}.}
\label{fig:p_vs_kil}
\end{figure}

Three sets of measured ion loss rates as a function of the photodetachment laser power are plotted in Fig.\ \ref{fig:p_vs_kil} for three different trap temperatures. Examples for the observed individual exponential decays are shown in the insets. For each trap temperature the rotational or collision temperature of the OH$^-$ interacting with helium has been determined separately, as explained above. The loss rates show a clear non-linear dependence on the power, indicative of the quenching dynamics of the first excited rotational state. To retrieve the inelastic collision rates we fit Eq.\ (\ref{eq:fit}) to the experimental data (solid line). The resulting parameters $\gamma_{10}$ are presented for all eight measurement sets in table \ref{tab:summary} together with their fitted statistical accuracy.

In addition to the statistical accuracy we have estimated the influence of the uncertainty of the measured rotational temperature on the inelastic collision rates: The measured power dependent ion loss rate was fitted assuming different rotational temperatures, namely $T_{rot}$, $T_{rot} + \Delta T_{rot}$ and $T_{rot} - \Delta T_{rot}$. The resulting deviations are deduced to be $\varepsilon_{\gamma}^{\Delta T} = \frac{1}{2} \cdot |\gamma_{10}(T_{rot} + \Delta T_{rot}) - \gamma_{10}(T_{rot} - \Delta T_{rot})| $. The values are comparable to the statistical accuracy (see Table \ref{tab:summary}).

\begin{figure}[tb]
\includegraphics[width=1.05\columnwidth]{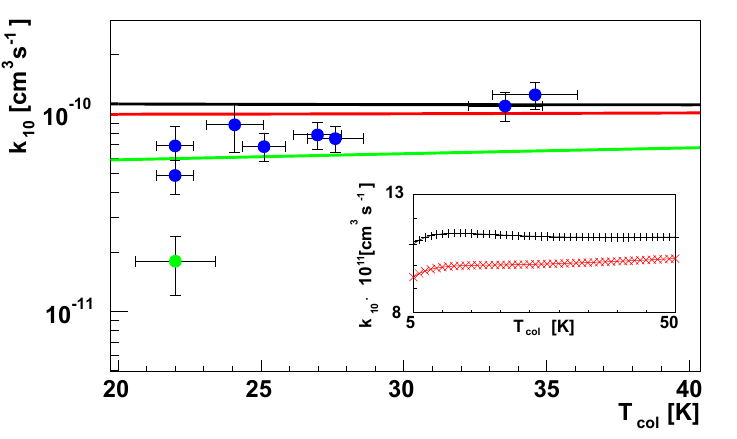}
\caption{
\label{fig:rate_vs_T}
$J=1\rightarrow 0$ inelastic rate coefficients as a function of the collision temperature (blue dots). The green dot shows the measurement from Ref.\ \cite{Hauser2015:np}. The black and red lines show the computational results for the $r_0$ and $v_0$ PESs, respectively, and the green line the theoretical calculation for the earlier PES \cite{Gonzalez-Sanchez2006:jpb,Hauser2015:np}.}
\end{figure}

The measured inelastic collision rates $\gamma_{10}$ depend linearly on the helium buffer gas density \cite{Hauser2015:np}. The rate coefficients $k_{10}$, obtained by division through this density, is therefore subject to a larger uncertainty due to the systematic accuracy of the density measurement. Several sources add up to a relative uncertainty of the density determination of 15\,\% (see supplementary material). The resulting rate coefficients are provided in Table \ref{tab:summary} and in Fig.\ \ref{fig:rate_vs_T}, where they are compared with calculations.

\begin{table}[tb]
\centering
 \caption{Results of the inelastic rate fits at different trap temperatures and buffer gas densities.}
\begin{tabular}{cccccc}
\hline
\hline
$T_{trap}$\,[K] & $T_{rot}$\,[K] & 
$\gamma_{10}$\,[s$^{-1}$] & $\epsilon_\gamma^{\Delta T}$\,[s$^{-1}$] & 
$\rho$\,[cm$^{-3}$] & $k_{10}$\,[$10^{-11}\frac{cm^3}{s}$]\\
\hline
9  & 22.0(7)  &  16.7(20) & 1.3 & 3.4$\cdot 10^{11}$ & 4.9(10)  \\
9  & 22.0(7)  &  16.5(31) & 1.2 & 2.4$\cdot 10^{11}$ & 6.9(18) \\
15 & 25.1(7)  &  18.1(16) & 1.2 & 2.6$\cdot 10^{11}$ & 6.9(11) \\
15 & 24.1(10) &  16.3(30) & 1.5 & 1.8$\cdot 10^{11}$ & 8.9(25) \\
20 & 27.0(8)  &  17.9(16) & 1.2 & 2.3$\cdot 10^{11}$ & 7.9(13) \\
20 & 27.6(10) &  12.0(9)  & 1.0 & 1.6$\cdot 10^{11}$ & 7.5(12) \\
30 & 33.5(13) &  20.5(18) & 1.6 & 1.9$\cdot 10^{11}$ & 11.0(18) \\
30 & 34.6(15) &  16.4(12) & 1.3 & 1.3$\cdot 10^{11}$ & 12.6(19) \\
\hline\hline
\end{tabular}
\label{tab:summary}
\end{table}

To precisely calculate the collisional quenching rates of excited rotational states, we have constructed two new potential energy surfaces (PESs), which we compare to an existing PES from our earlier work \cite{Gonzalez-Sanchez2006:jpb,Hauser2015:np}. The two new PESs are computed using coupled-cluster theory beginning with CCSD(T) at the complete basis set limit and adding the small contribution of electron correlation at the CCSDT(Q) level. For rotational relaxation the angular anisotropy of the surface is of particular importance. The first PES (denoted $r_0$) fixes the OH bond distance at its vibrationally averaged ground state distance. The second PES (denoted $v_0$) was constructed by averaging the electronic energies over the diatomic ground state vibrational probability density. The third PES was
obtained using MP4 theory and the rigid rotor geometry of the anion \cite{Gonzalez-Sanchez2006:jpb,Hauser2015:np}. Details on the electronic structure calculations, fitting  procedures and properties of PESs are provided in the supplementary material.

Scattering theory for diatomic molecules is well documented in the literature \cite{Gianturco1979springer,manolopoulos1986improved,hutson2007molecular}. The time independent collision dynamics calculations where done with the MOLSCAT code \cite{hutson2019molscat} for the first two PESs and with the ASPIN code \cite{Lopez-Duran2008:cpc} for the third PES. We verified that, on the same potential, calculations using the MOLSCAT code or the ASPIN code produce the same cross-sections. The calculations were performed with the formally exact close coupling method in the 10$^{-6}$ -- 10$^{3}$ cm$^{-1}$ energy range (for details see supplementary material).

\begin{figure}[tb]
\includegraphics[width=\columnwidth]{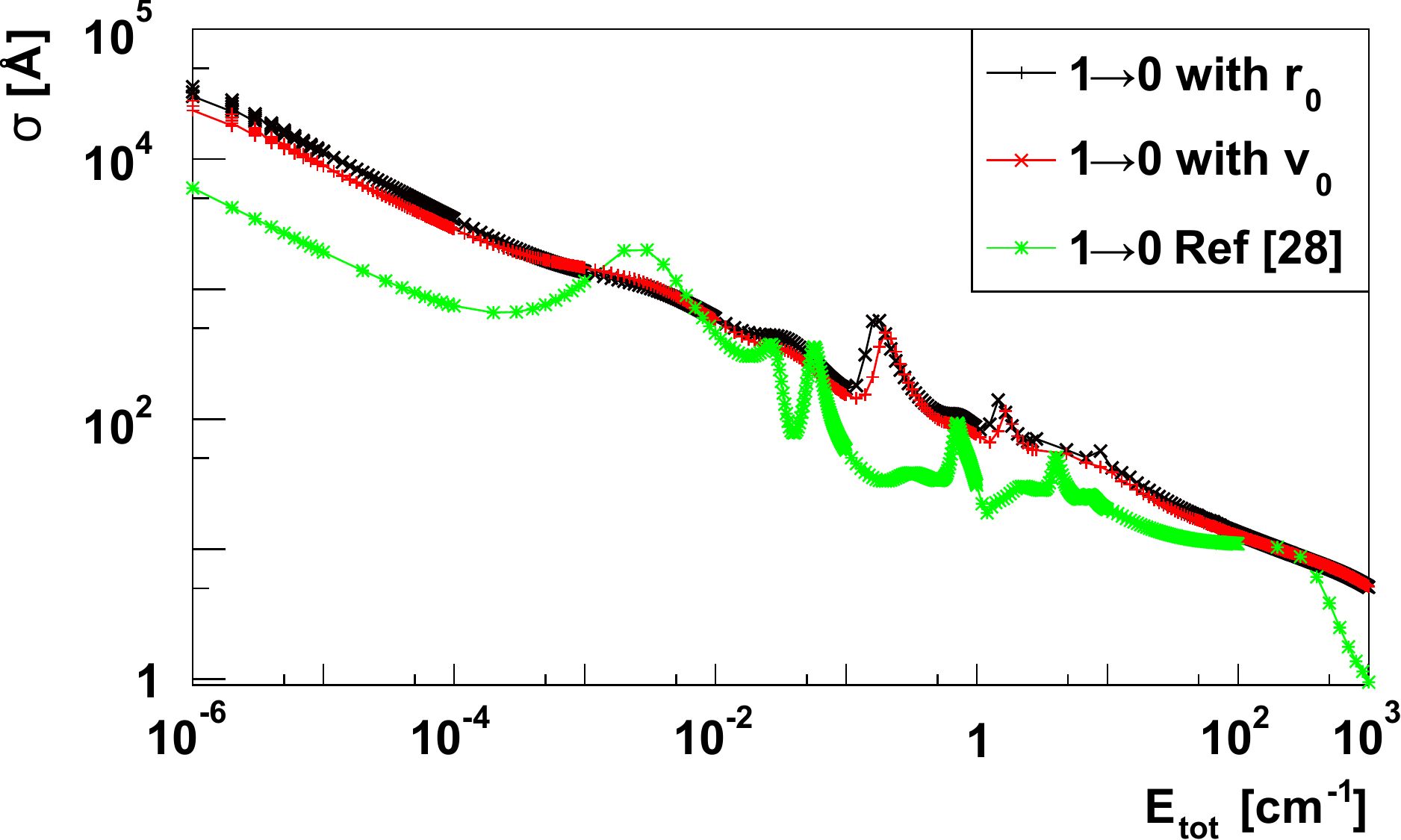}
\includegraphics[width=\columnwidth]{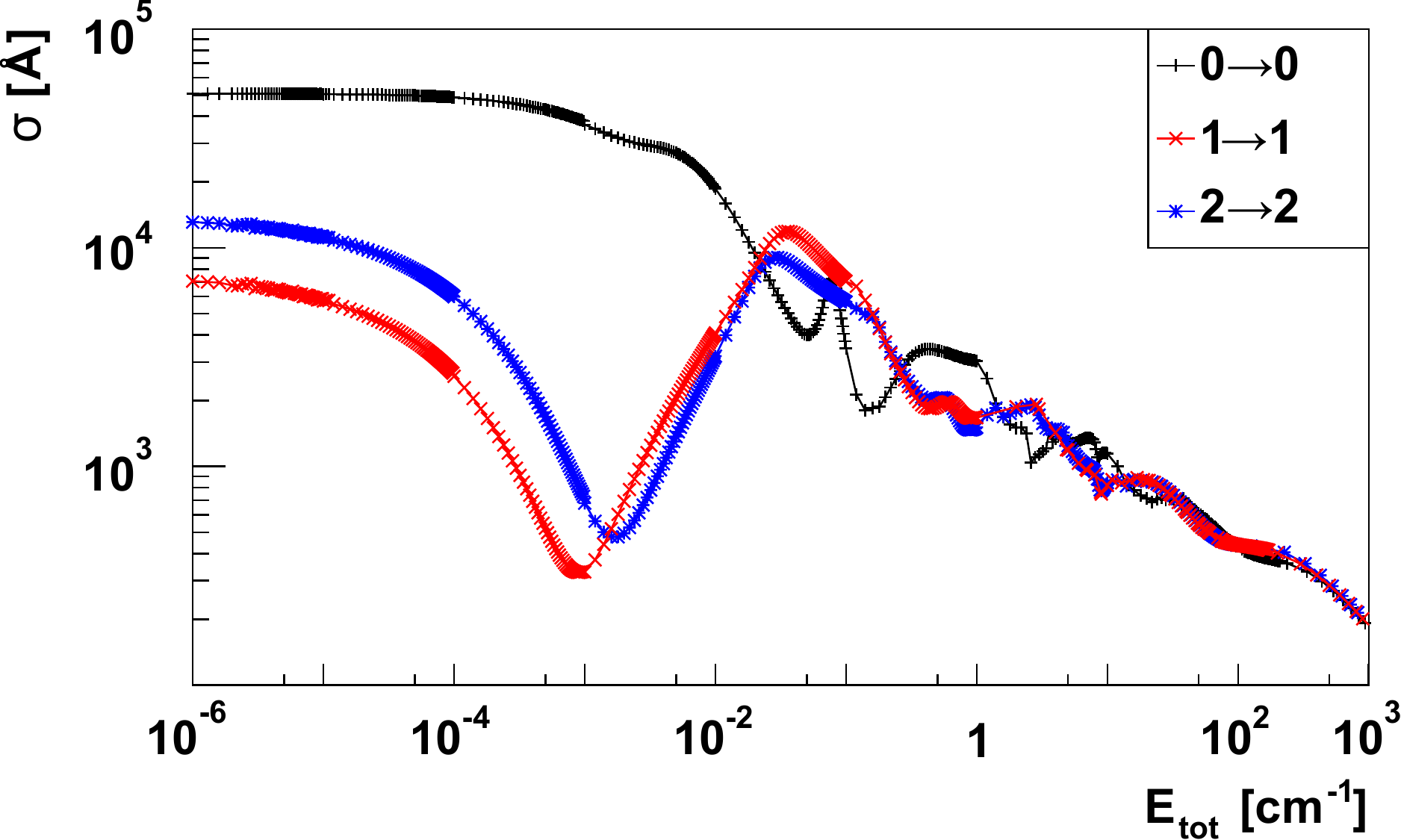}
\caption{Upper panel: Inelastic cross sections for OH$^-$ + He $J=1\rightarrow 0$ collisions. The results obtained from the  $r_0$ and $v_0$ PESs are compared with the result from the PES from Ref.\ \cite{Gonzalez-Sanchez2006:jpb,Hauser2015:np}. Lower panel: Elastic cross sections for OH$^-$ + He collisions in $J$=0, 1 and 2 for the $v_0$ PES.}
\label{fig:el-inxs}
\end{figure}

The obtained inelastic cross-sections of the $J=1\rightarrow 0$ transition are depicted in Fig.\ \ref{fig:el-inxs} (upper panel) for the three surfaces discussed here. As one can see, the cross sections from the $v_0$ PES are a little smaller than those from the $r_0$ PES, while both deviate significantly from the result from the earlier PES. Specifically, the two former cross sections are larger than the latter, in particular in the low-energy regime. The third cross section also shows stronger modifications due to resonance features, which can be linked to differences in the anisotropic parts of the PESs. In addition, Fig.\ \ref{fig:el-inxs} (lower panel) shows elastic cross sections obtained from the $v_0$ PES, for the lowest three rotational states of OH$^-$. These elastic cross sections are about one order of magnitude larger than the inelastic cross section, which is evidence for a more efficient cooling of the translation degrees of freedom compared to rotation. The same type of results have been obtained using the earlier PES \cite{Gonzalez-Sanchez2006:jpb,Hauser2015:np}.

In Fig.\ \ref{fig:rate_vs_T} the reaction rate coefficients, obtained from thermal averaging of the calculated cross sections with the different PESs, are plotted as a function of the collision temperature and compared to the experimental rate coefficients. In addition, we also present the experimental rate coefficient determined in Ref.\ \cite{Hauser2015:np}. That value is found to be smaller by a factor of three, which corresponds to about 2$\sigma$ deviation. While the statistical probability for such a deviation is still finite, we also investigated possible systematic sources of this deviation and found the long-term drift of a calibration factor for the density determination in the previous experiment as a possible source.

Overall the agreement between experiment and theory in Fig.\ \ref{fig:rate_vs_T} is very favorable. Interestingly, however, all three PESs predict slightly different values for the rate coefficients, differing by almost a factor of two. For the first two PES, the values are significantly larger than the experimental values at low tempreatures up to 30\,K, while they agree well for the two measurements around 35\,K. The theoretical rates from the earlier PES agree quite well with experiment at the lower temperatures, but are markedly smaller than the experimental values obtained around 35\,K. The vibrationally averaged PES produces reduced rate coefficients by about 10\% compared to the first PES with fixed bond distance, but this does not markedly change the comparison with the experiment.

The trend of a clear increase of the rate coefficients with temperature is observed in the experiment. This is somewhat captured by the calculations using the earlier PES \cite{Gonzalez-Sanchez2006:jpb,Hauser2015:np}, but it is not reproduced by the new calculations using the $v_0$ and $r_0$ PESs. This shows that rather subtle differences of the interaction potentials lead to observable differences in scattering rates. More accurate potential surface calculations or the explicit inclusion of vibrational excitation in three-dimensional scattering calculations may be needed to resolve this. Additionally, experiments at higher temperatures are desired to shed more light on the true shape of this trend. Ideally, then the next higher rotational state $J=2$ should also be measured and compared with calculations.

In summary, rotational state-selective removal of ions by threshold photodetachment has been applied to measure temperature-dependent inelastic collision rate coefficients of the OH$^-$ anion with helium. A precise analysis of the error budget was carried out including all systematic and statistical error. The experimental data are compared with first-principle theoretical calculations. The present experimental accuracy allowed us to test different rates calculated with three different potential energy surfaces. While in principle a good agreement was found for all the three surfaces, the calculations produced rate coefficients with two different temperature dependences that were both weaker than the measured temperature-dependence. This suggests that more work on the potential surface calculations and comparison of rate coefficients over a broader temperature range are needed in order to gain precise insight in the quantum effects at play.

We expect that the presented progress will also stimulate studies of more complex systems, such as polyatomic and open shell molecules. Recently we already analyzed the collisional quenching kinetics of NH$_2^-$ in helium buffer gas \cite{Gianturco2018:fd}, which can be extended to extract inelastic rate coefficients. For the OH$^+(^3\Sigma^-)$ cation, which has been detected in the Orion bar and other interstellar environments, rotational state-selective photodissociation may be used to measure inelastic scattering and test recent quantum inelastic scattering calculations \cite{Gonzalez-Sanchez2018:cpc}.

\begin{acknowledgments}
This work has been supported by the European Research Council under ERC Grant Agreement No. 279898 and by the Austrian Science Fund (FWF) through Project No. P29558-N36. E.S.E.\ acknowledges support from the Fonds National de la Recherche Luxembourg (Grant No. 6019121). R.D.\ is supported by the U.S. Department of Energy, Office of Science, Office of Basic Energy Sciences (Award DE-SC0019740).
\end{acknowledgments}

\end{document}